\documentclass[10pt]{article}

\usepackage{amsmath}
\usepackage{amssymb}
\usepackage{graphicx}
\usepackage{cite}
\usepackage{color} 

\usepackage[labelfont=bf,labelsep=period,justification=raggedright]{caption}
\makeatletter
\renewcommand{\@biblabel}[1]{\quad#1.}
\makeatother

\date{}

\newcommand{\avg}[1]{\langle{#1}\rangle}
\newcommand{\Avg}[1]{\left\langle{#1}\right\rangle}
 
\def\adjMatA{{\bf A}}
\def\elemAdjMatA{A}
\def\adjMatB{{\bf B}}
\def\elemAdjMatB{{B}}

\def\dampingA{{\alpha_{A}}}
\def\dampingB{{\alpha_{B}}} 

\def\degreeBIn{k_{\rm in}^{(B)}}
\def\degreeBOut{k_{\rm out}^{(B)}}
\def\inDegree{k^{\rm in}}
\def\outDegree{k^{\rm out}}
\def\degreeClass{{\bf k}}
\def\expo{\eta}

\usepackage[labelformat=simple]{subfig}

\begin{document}

\begin{flushleft}
{\Large 
\textbf{Multiplex PageRank}
}
\\
Arda Halu $^{1}$,
Ra\'ul J. Mondrag\'on$^{2}$, 
Pietro Panzarasa$^{3}$
Ginestra Bianconi $^{4,\ast}$,
\\
\bf{1} Department of Physics, Northeastern University, Boston, MA, USA
\\
\bf{2} School of Electronic Engineering and Computer Science, Queen Mary University of London, London, United Kingdom
\\
\bf{3} School of Business and Management, Queen Mary University of London, London, United Kingdom
\\
\bf{4} School of Mathematical Sciences,  Queen Mary University of London, London, United Kingdom
\\
$\ast$ E-mail: ginestra.bianconi@gmail.com
\end{flushleft}

\section*{Abstract} 
Many complex systems can be described as multiplex networks in which the same nodes can interact with one another in different layers, thus forming a set of interacting and co-evolving networks. Examples of such multiplex systems are social networks where people are involved in different types of relationships and interact through various forms of communication media. The ranking of nodes in multiplex networks is one of the most pressing and challenging tasks that research on complex networks is currently facing. When pairs of nodes can be connected through multiple links and in multiple layers, the ranking of nodes should necessarily reflect the importance of nodes in one layer as well as their importance in other interdependent layers. In this paper, we draw on the idea of biased random walks to define the Multiplex PageRank centrality measure in which the effects of the interplay between networks on the centrality of nodes are directly taken into account. In particular, depending on the intensity of the interaction between layers, we define the Additive, Multiplicative, Combined, and Neutral versions of Multiplex PageRank, and show how each version reflects the extent to which the importance of a node in one layer affects the importance the node can gain in another layer. We discuss these measures and apply them to an online multiplex social network. Findings indicate that taking the multiplex nature of the network into account helps uncover the emergence of rankings of nodes that differ from the rankings obtained from one single layer. Results provide support in favor of the salience of multiplex centrality measures, like Multiplex PageRank, for assessing the prominence of nodes embedded in multiple interacting networks, and for shedding a new light on structural properties that would otherwise remain undetected if each of the interacting networks were analyzed in isolation. 

\section*{Introduction}
Despite recent advances \cite{RMP,Newman_rev,Dynamics,Boccaletti2006} in the analysis of complex networks, a number of areas of investigation concerned with the description, prediction, and control of the dynamics of a variety of systems, including weather networks\cite{Kurths}, social networks\cite{Thurner}, and the brain\cite{Bullmore}, still remain largely unexplored. A large number of these systems cannot be properly understood unless they are regarded as components of higher-level systems in which various networks are connected with one another through a complex pattern of interdependencies~\cite{Boccaletti,Havlin1,Mucha,Thurner}. These higher-level systems can thus be seen as networks in which the interacting nodes are networks themselves that are characterized by their own structure and function, and co-evolve over time according to various patterns. Examples of such networks of networks include multimodal transportation networks, social networks, climatic systems, economic markets, energy--supply networks, and the human brain. For instance, the same individuals, groups and organizations can play different roles within a social relationship or can be linked through different types of social relationships (e.g., family relationships, acquaintanceship, friendship, and professional collaboration)~\cite{Kapferer1969,Lomi2006,Uzzi1996,Verbrugge1979,Zelizer2005}, can have different affiliations \cite{Lee2011,Wheeldon1969}, and can communicate with one another using different  technologies, such as mobile phone, chat, e-mail, or video conferences \cite{Baym2004,Huang2009}. Each of these roles, relationships, affiliations, and communication technologies can in turn be associated with a different social network in which links between nodes refer to a distinct form of social interaction between the connected nodes. The same people, groups or organizations that interact in many different ways can thus be represented as the nodes of multiple co-evolving social networks that are themselves connected with one another as the various forms of social interaction affect one another over time~\cite{Gibbons2004,Ingram2008,Lincoln1979,Maggioni2013}. Despite the ubiquity of these co-evolving and interconnected forms of interaction, so far network scientists have focused primarily on network datasets that contain only one type of social relation, thus neglecting the complexity of the connections between the various networks in which the same people interact. To uncover the nature and full breadth of social interaction, a special emphasis should be placed precisely on the structure and dynamics of the network of interacting social networks. 

The system in which the same nodes belong to multiple interacting and co-evolving networks is typically referred to as a multiplex network or multigraph~\cite{Flament,Wasserman}. In recent literature, there has been an upsurge of interest in multiplex networks. In particular, scholars have concentrated on the structural properties~\cite{Boccaletti,Maggioni2013,Mucha,Rank2010,Thurner} and the antecedents \cite{Ferriani2012,Shipilov2012} of these networks, have shed light on diffusion processes~\cite{Diffusion,Arenas2}, cooperation \cite{Gimeno,Games}, exchange relations \cite{Kuwabara2011,Kuwabara2010}, percolation phase transitions~\cite{Havlin1,JSTAT}, cascades\cite{Leicht}, epidemic spreading \cite{Boguna}, and election processes \cite{Arda_2013} occurring on them, and have developed modeling frameworks~\cite{Growth1,Growth2,Bianconi_2013} and game-theoretic perspectives~\cite{Games}. Among the structural properties of multiplex networks that scholars have only recently begun to address \cite{Boccaletti_ranking,Sheri}, a crucial role is played by the centrality of nodes. In a multiplex network, the importance of a node depends on the connectivity patterns within and across the different layers of the network. For univariate networks in which no more than one link can connect the same pairs of nodes, a number of measures are available for assessing the importance of nodes. Over recent years these measures have become increasingly popular and salient for a variety of empirical domains. Among these measures, in this paper we concentrate our attention on PageRank, a  centrality measure that has been successfully used not only for ranking web pages~\cite{brin1998anatomy}, but also for ranking scientists in citation networks~\cite{liu2005co} or species in food webs~\cite{allesina2009googling}. While PageRank was originally proposed as a centrality measure for univariate networks~\cite{brin1998anatomy}, its extension to multiplex networks remains largely unexplored. In particular, when the same pairs of nodes can be connected through multiple links co-evolving in multiple layers, a non-trivial problem is concerned with how to extend PageRank so as to capture the degree to which the ranking of nodes in one layer can affect, and be affected by, the ranking of the same nodes in other layers. This paper attempts to address this problem by proposing a generalization of PageRank to the case of multiplex networks.
  
To evaluate the relative popularity of a node in a network, PageRank centrality draws on the idea of a web surfer that visits different parts of the WWW at random. The random walker follows two strategies: the first is to jump to a node selected uniformly at random; the second is to jump, still randomly, to one of the walker's neighbors. The popularity of a node is a function of the frequency with which the random walker visits the node. This frequency is then compared with the frequencies associated with all other nodes in the network. The ranking of nodes obtained according to these frequencies of being visited is precisely the ordering produced by the PageRank centrality measure, and reflects the relative popularity that each node has across the whole network. 
 
To extend the PageRank centrality measure to the case of multiplex networks, we assume that the centrality a node has in one layer affects the centrality the node can obtain in another layer. This interplay between layers has a two-fold nature. First, the importance of a node in one layer may simply contribute to an increase in the node's importance in another layer. Second, a node's importance in one layer can amplify the node's ability to derive benefits from the importance of other nodes that point to it in another layer. Alternatively, from the perspective of a biased random walk on complex networks, Multiplex PageRank can be described in terms of the bias that one layer exerts on the random jumps that a surfer makes in another layer \cite{Vito,Holyst}. In this paper, we identify four versions of the Multiplex PageRank centrality measure, depending on how layers affect each other or, alternatively, exert a bias upon the random jump. First, if the bias lies in the jump the random walker makes to any other node in the network, we obtain a measure that we call \textit{Additive} Multiplex PageRank. Second, if  the bias is exerted upon the jump the walker makes to any of its neighboring nodes, we obtain the \textit{Multiplicative} Multiplex PageRank. The third variant is motivated by the fact that it is possible to have a bias in both jumps. In this case, we obtain the \textit{Combined} Multiplex PageRank. Finally, the \textit{Neutral} Multiplex PageRank refers to the case in which there is no bias in either jump, and thus the measure reduces to PageRank based on one single layer. For uncorrelated networks, we show how these centrality measures are correlated with the structural properties of the networks.

To clarify the meaning of the four versions of Multiplex PageRank, we apply these measures to a multiplex network formed by the juxtaposition of two networks, and show that the centrality of a node in one network depends on the centrality of the same node in the other network. Our application is concerned with online communication and is based on a multiplex network in which the same users can interact by sending instant messages to one another and by posting messages to a forum. As users can send messages directly to one another and at the same time participate in discussion groups within a forum, they can be regarded as embedded in two related online social networks. Our results show that the Multiplicative Multiplex PageRank of users displays a broad distribution, and is thus able to capture the emergence of high-ranked nodes, unlike what can be obtained through the application of the PageRank centrality measure to a single network.

\section*{Results}

Introduced as a centrality measure for assessing the ``importance'' of web pages, the PageRank $x_i$ of a node $i$ in a network with $N$ nodes is defined as~\cite{brin1998anatomy} 
\begin{equation}
x_i = \dampingA \sum\limits_{j} \elemAdjMatA_{ij} \frac{x_j}{g_j} + (1 - \dampingA)\frac{1}{N},
\label{Pagerank}
\end{equation}
where $\elemAdjMatA_{ij}$ are the elements of the adjacency matrix that are equal to one if node $j$ points to node $i$ and zero otherwise, $g_j=\max(1,\outDegree_j)=\max(1,\sum_{r}\elemAdjMatA_{rj})$, and $\dampingA>0$ is called the damping factor. PageRank can be interpreted as the stationary distribution of a random walk with additional random jumps. A random walker on site $j$ jumps to one of $j$'s $\outDegree_j$ out-neighbors with probability $\dampingA$, and to any other site chosen uniformly at random with probability $1-\dampingA$. The PageRank of a node is large to the extent that many other nodes point to it. The PageRank of a node is therefore expected to increase as a function of the node's in-degree, and indeed in \cite{Fortunato1,Fortunato2} it was shown that, for uncorrelated networks, the PageRank of nodes can be approximated by their in-degree. 
If nodes in uncorrelated networks are grouped into classes depending on their extended degrees $\degreeClass=(\inDegree, \outDegree)$, then the average PageRank for nodes of degree-class $\degreeClass$ is
\begin{equation}
\label{eq:PageRankSingle}
\overline{x}(\degreeClass)=\dampingA \frac{\inDegree}{\avg{\inDegree}N}+(1-\dampingA)\frac{1}{N},
\end{equation}
where the symbol $\avg{\dots}$ indicates the average over  the $N$ nodes of the network.

PageRank was originally proposed for ranking web pages in response to text queries, and for this reason it was formalized as a centrality measure for directed networks~\cite{brin1998anatomy}. It is, however, possible to extend the original definition to the case of undirected networks. For these networks, PageRank is 
\begin{equation}
x_i = \dampingA \sum_j \elemAdjMatA_{ij} \frac{x_j}{g_j} + (1 - \dampingA)\frac{1}{N},
\label{Pagerankundi}
\end{equation}
where $g_j=\max(1,k_j)$ and $k_j$ is the degree of node $j$. For undirected networks, the average PageRank $\overline{x}(k)$ of a node with degree $k$ is given by
\begin{equation}
\overline{x}({ k})=\dampingA \frac{k}{\avg{k}N}+(1-\dampingA)\frac{1}{N}.
\end{equation}

Compared to univariate networks, multiplex networks enable nodes to be connected with one another through more than one type of links, and as such offer a richer and more detailed backdrop against which the structural position of nodes can be assessed. The extension of PageRank to multiplex networks is therefore expected to shed light on novel ways for measuring the importance of nodes that capture their embeddedness in multiple interrelated relations. A ranking of nodes can thus be obtained that is likely to differ from the one originating simply from the position of nodes in one single network.  

Here we offer a generalization and extension of the PageRank measure that can be applied to any multiplex network dataset. The assumption underlying our proposed measure is that the centrality of a node in one network can be affected by the centrality of the same node in another network. For the sake of simplicity, we consider the case in which the multiplex network is organized into two layers: network A and network B. Our analysis can easily be generalized to multiplex networks with more than two layers (See Materials and Methods). We indicate with $\elemAdjMatA_{ij}$ the elements of the adjacency matrix of network A, and with $\elemAdjMatB_{ij}$ the elements of the adjacency matrix of network B. For network A, we evaluate PageRank ${\bf x} = \{x_1, \ldots, x_N \}$ using Eq.~(\ref{Pagerank}) with the parameter $\dampingA>0$. We then express the {\it Multiplex PageRank centrality} ${\bf X}=\{ X_1, \ldots , X_N \}$ of the nodes in network B with respect to PageRank ${\bf x}$. 

Formally, we define the Multiplex PageRank centrality $X_i$ of node $i$ as
\begin{equation}
X_i = \dampingB \sum\limits_j x_i^{\beta}\elemAdjMatB_{ij} \frac{X_j}{G_j} + (1 - \dampingB)\frac{x_i^{\gamma}}{N\avg{x^{\gamma}}},
\label{PageRankX}
\end{equation}
where { $G_j=\sum_{r} \elemAdjMatB_{rj}x_r^{\beta}+\delta(0,\sum_{r} \elemAdjMatB_{rj}x_r^{\beta})$, $\delta(a,b)$ is the Kronecker delta, }$\dampingB>0$ is small enough to guarantee that the relation can be satisfied, and the exponents $\beta$ and $\gamma$ are both greater than or equal to zero. The first term in Eq.~(\ref{PageRankX}) refers to the contribution to node $i$'s centrality that derives from the centrality of the nodes pointing to $i$ in network B. Like with the ordinary PageRank measure, this contribution is inversely proportional to the out-degree of node $i$'s in-neighbors. However, unlike the ordinary measure, Eq.~(\ref{PageRankX}) enables this contribution to be also affected by the centrality that both node $i$ and its in-neighbors in network B have in network A. This interplay between the two networks has a two-fold effect on a node's centrality. First, the extent to which node $i$ can derive some advantage from the centrality of its in-neighbors in network B becomes more significant as the centrality of $i$ in network A becomes larger. The more prominent a node is in one layer, the more likely it is that the node can attract and gain benefit from other important nodes in another layer. Second, the contribution of each in-neighbor $j$ to $i$'s centrality in network B is discounted by dividing $j$'s centrality by the sum of the centralities that $j$'s out-neighbors in network B have in network A. In other words, the benefits node $i$ can derive from the centrality of any in-neighbor $j$ in network B are diluted to the extent that $j$ in network B points to many other nodes that are associated with high centrality in network A. An important node in one layer can attract important nodes in a different layer, but the benefit that can be gained in so doing are mitigated if there are many other nodes that have a similar capacity of attraction. 

The second term in Eq.~(\ref{PageRankX}) reflects the contribution to node $i$'s centrality in network B that derives from $i$'s centrality in network A. By adding this second term, nodes that are not able to attract important neighbors in network B, can still derive some advantage simply by being central in network A. In the extreme case, a node with a zero in-degree in network B can still be associated with a non-zero value of centrality if the node has a non-zero centrality in network A. The assumption underlying this component of centrality is that the importance of a node in one layer is positively affected by the importance that the same node has in another layer, regardless of the node's capacity to attract other important nodes in the former layer.

Alternatively, Multiplex PageRank can also be regarded as the stationary distribution of a random walk with additional biased jumps. With probability $\dampingB$, a random walker on site $j$ jumps to site $i$, one of $j$'s $\outDegree_j$ out--neighbors selected with probability proportional to $x_i^{\beta}$, and with probability $1-\dampingB$ jumps to site $i$ chosen with probability proportional to $x_i^{\gamma}$. 

In what follows, we identify four important limiting cases of the Multiplex PageRank measure:
\begin{itemize}
\item {\em Additive Multiplex PageRank
($\beta=0$, $\gamma=1$)}:
\begin{equation}
X_i = \alpha_B \sum\limits_j B_{ij} \frac{X_j}{G_j} + (1 - \alpha_B)\frac{x_i}{N\avg{x}},
\label{additive}
\end{equation}
where $G_j=\max(1,\sum_{r}B_{rj})$. This refers to the case in which the effect of network A on network B is exerted simply by ``adding'' some value to the centrality the nodes have in network B in proportion to the centrality they have in network A. Here the interplay between networks does not imply that the importance a node has in one network affects the node's ability to derive benefits from important nodes in another network. Simply, being central in network A enables a node to gain more centrality in network B, regardless of the node's capacity to attract important others in network B. Recast in terms of a random walk, this version of Multiplex PageRank refers to the case of a biased random walk, where the bias lies in the random jump to any node in network B. In particular, nodes with high PageRank in network A are preferred over other nodes with low PageRank in the same network as the destination of the random jumps that the walker makes in network B. A similar version of this PageRank measure, in which the random jumps are biased according to some predetermined distribution called ``personalized vector", has already been suggested in the computer science literature \cite{Personalized}. In qualitative agreement with this version, here we propose to regard a node's PageRank in one layer as the node's ``personalized vector" in another layer.
\item{\em Multiplicative Multiplex PageRank ($\beta=1$, $\gamma=0$)}:
\begin{equation}
X_i = \alpha_B \sum\limits_j x_i B_{ij} \frac{X_j}{G_j} + (1 - \alpha_B)\frac{1}{N},
\label{multiplicative}
\end{equation}
where { $G_j=\sum_{r} \elemAdjMatB_{rj}x_r+\delta(0,\sum_{r} \elemAdjMatB_{rj}x_r)$.} This refers to the case in which the effect of network A on network B lies in ``multiplying'' the benefits that a node gains from the importance of its in-neighbors in network B by a factor that is proportional to the node's importance in network A. Thus, all benefits that can be obtained by being central in network A are contingent upon the connections that a node receives from important nodes in network B. The more important a node is in network A, the more value the node can extract from the connections received from important others in network B. Unlike the Additive version, the Multiplicative Multiplex PageRank does not enable a node to derive any added benefit in network B by simply being important in network A, regardless of the importance of the node's in-neighbors in network B. Alternatively, this version of the measure also refers to the case of a biased random walk, where the bias lies in the walker's choice of the out--neighbor as the destination of the jump. In particular, neighbors with high PageRank in network A are preferred over other neighbors with low PageRank in the same network.
\item{\em Combined Multiplex PageRank ($\beta=\gamma=1$)}:
\begin{equation}
X_i = \alpha_B \sum\limits_j x_i B_{ij} \frac{X_j}{G_j} + (1 - \alpha_B)\frac{x_i}{N\avg{x}},
\label{combined}
\end{equation}
where { $G_j=\sum_{r} \elemAdjMatB_{rj}x_r+\delta(0,\sum_{r} \elemAdjMatB_{rj}x_r)$.} This refers to the case in which the effect of network A on network B lies in ``combining'' the additive and multiplicative benefits a node in network B can gain by being central in network A. In this case, a node's high centrality in network A can boost its centrality in network B both in itself and at the same time by amplifying the node's ability to derive centrality from other important nodes. Alternatively, this version of Multiplex PageRank refers to the case in which both the destination of the random jump and the selection of the random walker's out-neighbor in network B are biased in that they favor nodes with high PageRank in network A over nodes with low PageRank in the same network.
\item{\em Neutral Multiplex PageRank} ($\beta=\gamma=0$):
\begin{equation}
X_i = \alpha_B \sum\limits_j B_{ij} \frac{X_j}{G_j} + (1 - \alpha_B)\frac{1}{N},
\label{neutral}
\end{equation}
where $G_j=\max(1,\sum_{r}B_{rj})$. This refers to the case in which there is no effect of network A upon network B, and thus Multiplex PageRank reduces to the PageRank based simply on network B in isolation.
\end{itemize}
Clearly these limiting cases can be generalized so as to be applied also to a multiplex network that combines two undirected networks or a directed network and an undirected one. Moreover, the above definitions can be further generalized so as to accommodate cases in which the rankings ${\bf x}$ are obtained using different centrality measures, such as the eigenvector centrality. 

Following \cite{Fortunato1, Fortunato2}, we performed a mean-field calculation of the average Multiplex PageRank $\overline{X}({\bf k}_B,x)$ of a node with degree ${\bf k}_B=(\inDegree_B,\outDegree_B)$ in network B and PageRank $x$ in network A. We define $\overline{X}({\bf k}_B,x)$ in the following way
\begin{equation}
\overline{X}({\bf k}_B,x)=\frac{1}{NP({\bf k}_B,x)}\sum_{i |{\bf k}_{B,i}={\bf k}_B,x_i=x}X_i,
\end{equation}
where $P({\bf k}_B,x)$ is the probability that a node has degree ${\bf k}_B=(k_B^{in},k_B^{out})$ in network B and PageRank $x$ in network A.
In particular, performing a mean-field calculation (see Materials and Methods) valid for an uncorrelated network B, we obtain 
\begin{equation}
\overline{X}({\bf k}_B,x)=\alpha_B x^{\beta}k^{in}_{B}\frac{1}{\avg{x^{\beta}k^{in}_B}N}+(1-\alpha_B)\frac{x^{\gamma}}{N\avg{x^{\gamma}}}.
\end{equation}
\begin{figure}
\begin{center}
\includegraphics[width=0.7\textwidth]{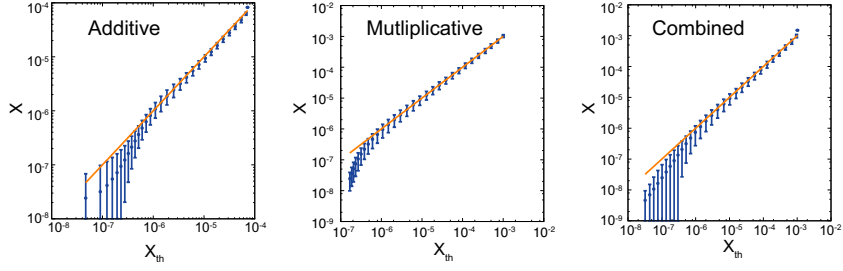}
\end{center}
\caption{\label{fig:PageRankOne} { {\bf Data  $X$ versus theory  $X_{th}$ for the Additive, Multiplicative, and Combined versions of PageRank.} In each of the three panels, the PageRank of the data is plotted against the corresponding value obtained through our theoretical approximation. Multiplex PageRank was evaluated using an iterative procedure with the standard values $\alpha_A=\alpha_B  = 0.85$. 
The accuracy of the algorithm was set at $10^{-11}$.} }
\end{figure}
To verify the validity of Multiplex PageRank, we generated a duplex network with $10^7$ nodes and $8\times10^7$ links in each layer. In both layers, the in- and out-degrees decay as a power law $k^\expo$, where $\expo_{A}^{\rm out} = 2.8$, $\expo_{A}^{\rm in}=2.1$ and $\expo_{B}^{\rm out} = 2.5$, $\expo_{B}^{\rm in}=2.5$. The dependence of the Additive, Multiplicative, and Combined versions of PageRank upon in-degree is shown in Fig.~\ref{fig:PageRankOne}. For small values of PageRank, the deviation from the diagonal is due to large fluctuations of PageRank in correspondence of small values of in-degree, as was also observed by Fortunato~et al.~\cite{Fortunato1}.

\section*{Discussion}
We apply the Multiplex PageRank measure to the multiplex network created from an online community at the University of California, Irvine~\cite{Panzarasa2009}. The multiplex network includes two layers. The first layer corresponds to a directed instant messaging (IM) network in which a directed link is established from one user to another if the former sends one or more online instant messages to the latter. The second layer is a bipartite network in which a link is established between a user and a discussion group of a forum when the former posts a message to the latter. While the IM network dataset covers the period from April 19 to October 26, 2004, the forum became active at a later time when users were already communicating through instant messages. The bipartite network thus covers a more restricted period than the IM network, from May 14 to October 26, 2004. The two networks also differ in the number of users: the total number of active users recorded for the IM network is $1,899$, of whom only $899$ posted at least one message in the forum. Moreover, users that were active in the forum created $552$ thematic groups, each aimed at the discussion of a specific topic. 
\begin{figure}
\begin{center}
\includegraphics[width=0.3\textwidth]{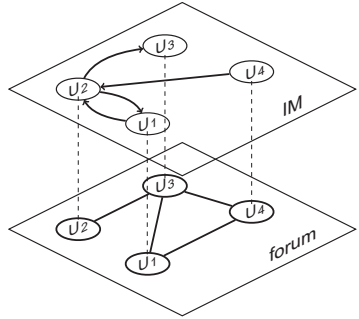}
\caption{\label{fig:Multiplex} {\bf Sketch of the multiplex online social network in which users communicate by exchanging instant messages and by posting messages to a forum.}}
\end{center}
\end{figure}

The analysis of the multiplex network covers the restricted observation period beginning on June 4, 2004, when both networks were operational and exhibited a fairy stable pattern of activity. At any specific day, and with a daily frequency, we constructed the instantaneous cumulative networks reflecting all the social interactions that took place in the three weeks' period ending on that day. Measurements thus create a time series with $124$ sample networks starting on June 25, 2004. The multiplex network can be represented by the juxtaposition of the two time-varying adjacency matrices $\adjMatA(t)$ and $\adjMatB(t)$ that describe the IM network and the one-mode projection of the bipartite forum network, respectively (see Fig.~\ref{fig:Multiplex}). In particular, the adjacency matrix $\adjMatA(t)$ describes directed links between users, i.e. $A_{ij}(t)=1$ if user $j$ sent at least one message to user $i$ in a given time window. For the forum network, the adjacency matrix $\adjMatB(t)$ describes an undirected and unweighted network between the users of the forum, where $B_{ij}(t)=1$ if both user $i$ and user $j$ posted at least one message to a common discussion group in a given time window. 

\begin{figure}[ht]
\centering
\includegraphics[width=0.7\textwidth]{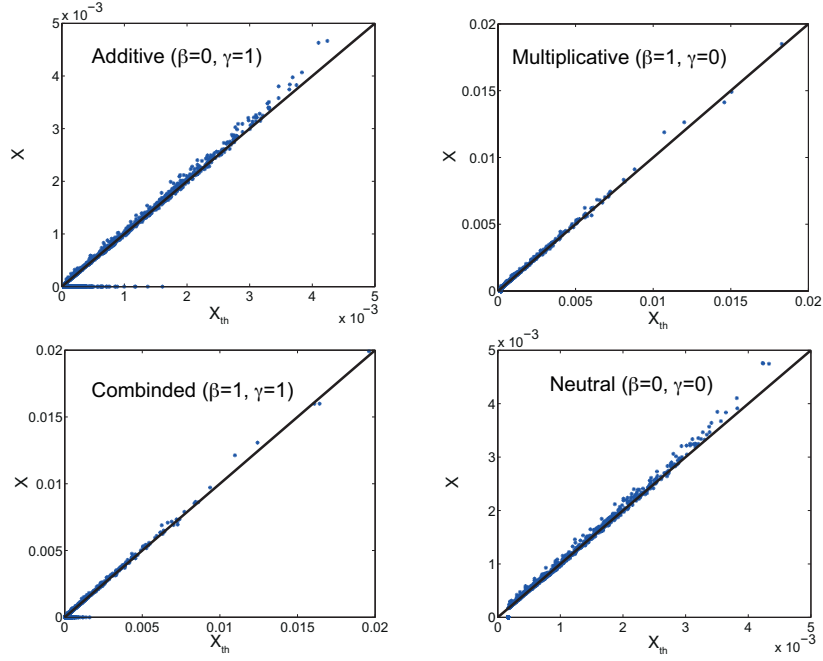}
\caption{\label{fig:approBipartite}{\bf The Additive, Multiplicative, Combined, and Neutral users' Multiplex PageRanks $X$ plotted against the mean field expectation $X_{th}$ (solid line) for the IM-forum multiplex network dataset.} The damping factors used for the IM and forum data are $\alpha_A = \alpha_B = 0.85$.}
\end{figure}
The application of Multiplex PageRank to online communication is motivated by the fact that users can enhance their ranking by engaging in multiple and interrelated ways of communication. In our specific case, users' prominence in the IM network (A) is likely to have an impact upon the prominence they gain by communicating and interacting in the forum network (B). To fully capture this intertwined nature of users' prominence, we begin by calculating each user's PageRank $x_i$ based on the IM network. The Multiplex PageRank $X_i$ of user $i$ in the forum is then obtained by expressing the PageRank user $i$ has in the forum as a function of the user's PageRank $x_i$ based on the IM network. Formally, the Multiplex PageRank of user $i$ in the forum at time $t$ is
\begin{eqnarray}
X_i(t)&=&\dampingB\sum_{j}[x_i(t)]^{\beta} \elemAdjMatB_{{ij}}(t)\frac{X_{j}(t)}{G_{j}(t)}+(1-\dampingB)\frac{[x_i(t)]^{\gamma}}{\avg{[x(t)]^{\gamma}}_B N_{F}(t)}\nonumber, \\
\end{eqnarray}
where $N_{F}(t)$ is the number of active users in the forum at time $t$, $\avg{\dots}_B$ denotes the average of $x(t)$ based only on the nodes that belong to network B at time $t$, and {$G_j(t)=\sum_{r} \elemAdjMatB_{rj}(t)[x_{r}(t)]^{\beta}+\delta(0,\sum_{r} \elemAdjMatB_{rj}(t)[x_{r}(t)]^{\beta})$.} In the above formula, the PageRank $x_i$ of node $i$ in the IM network at time $t$ is given by
\begin{eqnarray}
x_{i}(t)&=&\dampingB\sum_{j} \elemAdjMatA_{{ij}}(t)\frac{x_{j}(t)}{g_{j}(t)}+(1-\dampingB)\frac{1}{N},
\end{eqnarray}
with $g_{j}(t)=\max(1,k_j^{A,out}(t))$. For the IM-forum multiplex network, we compared the values of Multiplex PageRank with the theoretical expectations obtained in the case of an uncorrelated network. We found a very good agreement between the two sets of values (see Fig.~\ref{fig:approBipartite}).

A crucial issue affecting a large number of applications, ranging from the online sale of books to usage of Twitter tags, that rely on measures for ranking items is concerned with the stability of the rankings over time \cite{Ghoshal13}. To address this problem, here we investigate the stability of the top-ranked users in the forum, and compare the rankings obtained using the different proposed versions of Multiplex PageRank.

\begin{figure}[ht]
\centering
\includegraphics[width=0.7\textwidth]{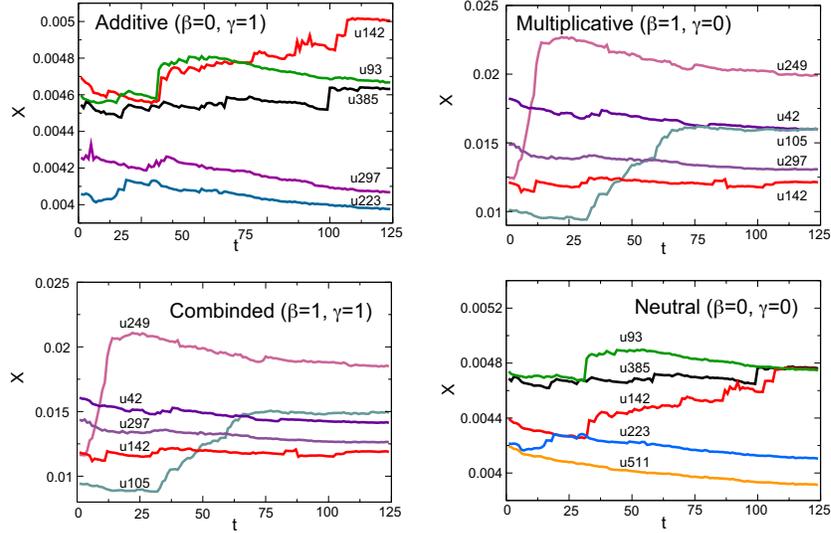}
\caption{{\bf The time evolution of the values of Additive, Multiplicative, Combined, and Neutral Multiplex PageRank $X$ for the $5$ top-ranked users.} The damping factors used for the IM and forum network data are $\alpha_A = \alpha_B = 0.85$. Each time step reflects the cumulative interactions in a three-week time window.}
\label{v2maxv2}
\end{figure}

In order to evaluate the stability of rankings in our dataset, we select the top five users with the highest Multiplex PageRank at the end of the whole observation period ($t = 124$), and track their evolution over time. In Fig.~\ref{v2maxv2} we plot the time evolution of the values of PageRank $X$ of the five users with the highest Additive, Multiplicative, Combined, and Neutral PageRanks. Users are ranked in decreasing order, from top to bottom. The figure indicates that the top five users with the highest values of Multiplicative Multiplex PageRank and Combined Multiplex PageRank are the same, and the top five users with the highest values of Additive and Neutral Multiplex PageRank are the same (with the exception of user 297 and user 511; note that user 297 has extremely high PageRank $x$ in the IM network). The Neutral Multiplex PageRank refers to the case in which $\beta=0$ and $\gamma=0$ in Eq.~(\ref{PageRankX}), and thus produces a ranking of users that coincides with the one obtained by taking into consideration only users' position in the forum.

When the importance of the users in the forum is adjusted to also reflect their position in the IM network simply by adding a bias in the random jump to any node, as occurs with the Additive Multiplex PageRank, the identity of the top five users in the resulting ranking does not change significantly. On the contrary, the introduction of a bias in the random walker's choice of the out-neighbor as the destination of the jump, as occurs with the Multiplicative Multiplex PageRank, is responsible for a substantial change in the ranking of users. In particular, the emergence of new top-ranked users when the Multiplicative Multiplex PageRank is adopted suggests that there are synergies between the activities of these users in the IM and forum networks. The way they communicate and rise to prominence in one network affects how they communicate and rise to prominence in the other network. Thus, taking into account the multiplex nature of the dataset helps unveil these synergies and the multi-faceted nature of users' prominence that would otherwise remain undetected if only one layer were investigated. Moreover, Fig.~\ref{v2maxv2} shows that the Combined Multiplex PageRank, by adding a bias both in the random jump to any node and in the walker's choice of the out-neighbor, does not produce any substantial change in the ordering of the top-ranked users with respect to the ranking that is obtained with the Multiplicative Multiplex PageRank.

\begin{figure}[ht]
\centering
\includegraphics[width=0.7\textwidth]{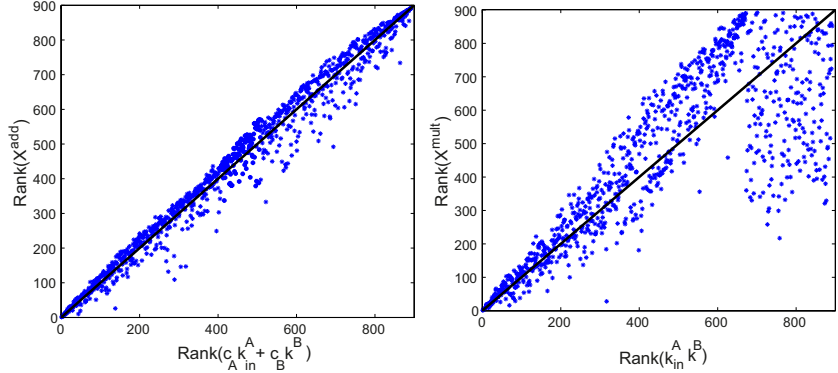}
\caption{\label{fig:simplerRanking} {\bf The estimated ranks of users according to the sum $(c_A k^A_{in} + c_B k^B)$ and product $(k^A_{in} k^B)$ of their in-degrees and degrees plotted against the values of their Additive and Multiplicative Multiplex PageRank, respectively.} In the figure $c_A = (1-\alpha_B)\alpha_A / (\langle k^A_{in}\rangle N \langle x\rangle_B)$ and $c_B = \alpha_B / \langle k^B\rangle$. Note that the node with rank $1$ is the most important node of the network, and therefore the Additive and Multiplicative Multiplex PageRanks of the most important nodes of the online social network are correlated, respectively, with a linear combination or the product of the users' in-degrees and degrees.}
\label{corr}
\end{figure}

In most applications, the use of PageRank for assessing the importance of nodes is aimed primarily at producing a ranking of nodes rather than associate each of them with a specific value of centrality. As in a variety of networks nodes' PageRank is closely related to their in-degree, especially for nodes with high in-degree, it has become common practice to use in-degree as a proxy for PageRank. In the case of the multiplex online social network, drawing on our theoretical framework, we tested the hypothesis that the Additive Multiplex PageRank correlates with a linear combination of nodes' in-degree in network A and their degree in network B, while the Multiplicative PageRank correlates with the product between nodes' in-degree in network A and their degree in network B. Fig.~\ref{fig:simplerRanking} does indeed provide support in favor of our hypothesis. Findings thus suggest that the Additive and Multiplicative versions of PageRank can be well approximated by the following two simple measures of centrality for multiplex networks: respectively, the linear combination of nodes' degrees in the different layers, and the multiplication of nodes' degrees in the different layers.

\begin{figure}[ht]
\begin{center}
\includegraphics[width=0.7\textwidth]{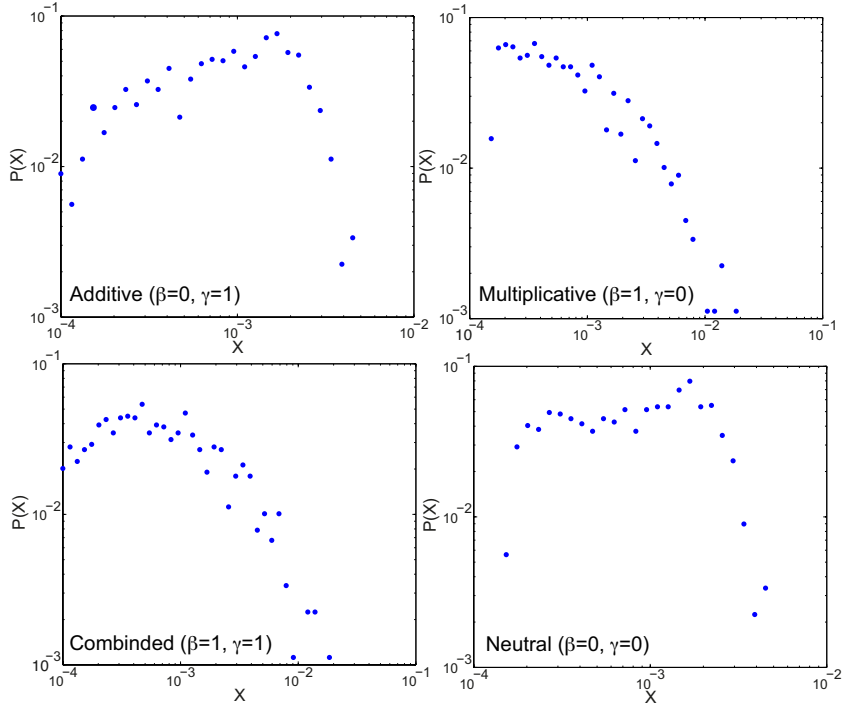}
\caption{{\bf The distribution $P(X)$ of the Additive, Multiplicative, Combined, and Neutral versions of Multiplex PageRank $X$ for users in the IM-forum multiplex network dataset.} The damping factors used for the IM and forum data are $\alpha_A = \alpha_B = 0.85$.}
\end{center}
\label{fig:dist}
\end{figure}

Finally, we found that in our dataset the distribution of Multiplex PageRank is broad, especially the one of the Multiplicative and Combined versions of Multiplex PageRank (See Figure $\ref{fig:dist}$), as is expected in the case of multiplex networks with positive correlations between degrees of nodes in the different layers.   

In conclusion, in this paper we introduced Multiplex PageRank, namely a centrality measure that can be used to identify and rank important nodes in multiplex networks. In particular, we defined four versions of this measure: the Additive, Multiplicative, Combined, and Neutral Multiplex PageRank.  We then analyzed how these measures correlate with the degree of the nodes in the different layers, both at the mean-field level and using data on an online social network. The empirical application of these measures to our dataset indicated that taking into consideration the multiplex nature of social interaction helps uncover the emergence of rankings of nodes and of structural properties that would otherwise remain undetected if only univariate single networks were investigated. 
\section*{Materials and Methods}
The dataset used in this paper consists of communication records of anonymized users that were randomly assigned an identification number. This ensures privacy protection and compliance with ethical guidelines. The dataset  can be obtained upon request from Pietro Panzarasa. The use of the dataset is free, provided the appropriate credit is given to the authors and a reference is made to this paper and to paper \cite{Panzarasa2009}.

\subsection*{Approximating Multiplex PageRank by the degree. Derivation of Eq. (11).}
In order to calculate the Multiplex PageRank $X_i$ of node $i$, we use an iterative procedure. PageRank ${\bf X}^n=\{X_1^n, \ldots ,X_N^n \}$ at time step $n$ can then be calculated from PageRank ${\bf X}^{n-1}$ at time step $n-1$ according to the recursive equation
\begin{equation}
X_i^{n}= \dampingB \sum_j x_i^{\beta} \elemAdjMatB_{ij} \frac{X_j^{n-1}}{G_j} + (1 - \dampingB)\frac{x_i^{\gamma}}{\avg{x^{\gamma}}N},
\label{iterative}
\end{equation}
where PageRank $X_i$ is given by $X_i=\lim_{n\to \infty} X_i^n$ and where $G_j=\sum_{r} \elemAdjMatB_{rj}x_r^{\beta}+\delta(0,\sum_{r} \elemAdjMatB_{rj}x_r^{\beta})$.

Following~\cite{Fortunato1, Fortunato2}, we divide the PageRank of nodes in network B into different classes, where two or more nodes belong to the same class if they have the same in- and out-degree and the same PageRank $x$. We define  
$\overline{X}({\bf k}^{(B)},x)$ as the average value of the PageRank of nodes in the degree class ${\bf k}^{(B)}=(\degreeBIn,\degreeBOut)$ in network B and with PageRank $x$ in network A
\begin{equation}
\overline{X}({\bf k}^{(B)},x)=\frac{1}{NP({\bf k}^{(B)},x)}\sum_{i | {\bf k}_{i}^{(B)}={\bf k}^{(B)},x_i=x}X_i,
\end{equation}
where $P({\bf k}^{(B)},x)$ is the probability that a node has degree ${\bf k}^{(B)}$ in network B and PageRank $x$ in network A.
Similarly, we define $\overline{X^n}({\bf k}^{(B)},x)$ from the iterative procedure
\begin{eqnarray}
\overline{X^n}({\bf k}^{(B)},x)&=&\frac{1}{NP({\bf k}^{(B)},x)}\sum_{i | {\bf k}_{i}^{(B)}={\bf k}^{(B)},x_i=x}X_i^{n}\nonumber\\&=&  \frac{\dampingB}{NP({\bf k}^{(B)},x)}\left(\sum_{i | {\bf k}_{i}^{(B)}={\bf k}^{(B)},x_i=x}\sum_j x_i^{\beta} \elemAdjMatB_{ij} \frac{X_j^{n-1}}{G_j}\right) + (1 - \dampingB)\frac{x^{\gamma}}{\avg{x^{\gamma}}N}.
\label{iterative2}
\end{eqnarray}
The term $G_j$ in the above equation can be approximated as
{\begin{equation}
G_j=\sum_{r} x_r^{\beta} \elemAdjMatB_{rj}+\delta(0,k_{j,{\rm out}}^{(B)})\simeq k_{{\rm out},j}^{(B)}\sum_{{\bf k'}^{(B)},x'}P^{(B)}_{\rm out}({\bf k'}^{(B)},{x'}|{\bf k}^{(B)}_j,{x_j}) (x')^{\beta}+\delta(0,k_{j,{\rm out}}^{(B)}),
\end{equation}}
where $P^{(B)}_{\rm out}({\bf k'}^{(B)},{x'}|{\bf k}^{(B)}_j,{x_j})$ is the probability of reaching a node with degree ${\bf k'}^{(B)}$ and PageRank $x'$ by following a link in network B from a node of degree ${\bf k}^{(B)}_j$ and PageRank $x_j$. If the nodes belonging to class $\{ {\bf k}^{(B)}_{j,},x_j \}$ are uncorrelated to the nodes of class $\{ {\bf k}^{(B)}, x \}$, then 
\begin{equation}
P^{(B)}_{\rm out} ({\bf k'}^{(B)},{x'}|{\bf k}^{(B)}_j,{x_j})= \frac{{k'}^{(B)}_{\rm in}}{\avg{\degreeBIn} }P({\bf k'}^{(B)},x'), 
\end{equation}
such that { 
\begin{equation}
G_j\simeq k^{(B)}_{{\rm out}, j}\frac{\avg{x^{\beta}\degreeBIn}}{\avg{\degreeBIn}}+\delta(0,k_{j,{\rm out}}^{(B)}).
\end{equation}}
Using the  approximation for $G_j$ we can express the sum in Eq.~(\ref{iterative2}) via a mean--field approximation, obtaining
\begin{eqnarray}
\sum_{i |{\bf k}^{(B)}_{i}={\bf k}^{(B)},x_i=x}\sum_j x_i^{\beta} \elemAdjMatB_{ij} \frac{X_j^{n-1}}{G_j}
&=&\sum_{\bf k''}^{(B)}\sum_{x''}\sum_{i |{\bf k}^{(B)}_{i}={\bf k}^{(B)},x_i=x}
\sum_{j|{\bf k}^{(B)}_{j}={\bf k''}, x_j=x''}
 x_i^{\beta} \elemAdjMatB_{ij} \frac{X_j^{n-1}}{k^{(B)}_{{\rm out},j}}\frac{\avg{k^{(B)}_{\rm in}}}{\avg{x^{\beta}k^{(B)}_{\rm in}}}\nonumber \\
&&\hspace*{-30mm}\simeq
\sum_{\bf k''}\sum_{x''}\sum_{i |{\bf k}^{(B)}_{i}={\bf k}^{(B)},x_i=x}\overline{X^{n-1}}({\bf k''}^{(B)},x'')  \frac{\avg{k^{(B)}_{\rm in}}}{\avg{x^{\beta}k^{(B)}_{\rm in}}} \frac{x_i^{\beta}}{(k'')^{(B)}_{\rm out}}\sum_{j|{\bf k}^{(B)}_{j}={\bf k''}x_j=x''}\elemAdjMatB_{ij} \nonumber \\
&&\hspace*{-30mm} \simeq \sum_{\bf k''}\sum_{x''}\sum_{i |{\bf k}^{(B)}_{i}={\bf k}^{(B)},x_i=x}\overline{X^{n-1}}({\bf k''}^{(B)},x'')   
\frac{\avg{k^{(B)}_{\rm in}}}{\avg{x^{\beta}k^{(B)}}}
 \frac{ x_i^{\beta} (k^{(B)})_{\rm in}}{(k'')_{\rm out}}P^{(B)}_{\rm in} ({\bf k''}^{(B)},x''  |{\bf k}^{(B)}, x)\nonumber,
\label{uno}
\end{eqnarray}
where we used the mean--field approximation 
\begin{equation}
\sum_{j | {\bf k}^{(B)}_{j}={\bf k''}, x_j=x''} X_j^{n-1}\elemAdjMatB_{ij} \simeq  \overline{X^{n-1}}({\bf k''}^{(B)},x'')
\sum_{j|{\bf k}^{(B)}_{j}={\bf k''}, x_j=x''} \elemAdjMatB_{ij}
\end{equation} 
and 
 $P^{(B)}_{\rm in} ({\bf k''}^{(B)},x''  |{\bf k}^{(B)}, x)$ is the probability that, by following an incoming link of a node with degree ${\bf k}^{(B)}$ and PageRank $x$ in network B, a predecessor of the node with degree ${\bf k''}^{(B)}$ and PageRank $x''$ can be reached.
In an uncorrelated network, this quantity is given by 
\begin{equation}
P^{(B)}_{\rm in} ({\bf k''}^{(B)},x''  |{\bf k}^{(B)}, x)=\frac{k''^{(B)}_{\rm out}}{\avg{k^{(B)}_{\rm in}}}P^{(B)}({\bf k''}^{(B)},x'').
\label{due}
\end{equation} 
Inserting Eqs.~$(\ref{uno})$ and $(\ref{due})$ in Eq.~$(\ref{iterative2})$, and taking the limit $n\to \infty$, we obtain
\begin{equation}
\overline{X}({\bf k}^{(B)},x)=\dampingB \frac{x^{\beta}k^{(B)}_{\rm in}}{\avg{x^{\beta}k^{(B)}_{\rm in}}N}+(1-\dampingB)\frac{x^{\gamma}}{N\avg{x^{\gamma}}},
\end{equation}
where the values of $x$ can be approximated from Eq.~(\ref{eq:PageRankSingle})
\begin{equation}
\overline{x}({\bf k}^{(A)})=\dampingA \frac{k^{(A)}_{\rm in}}{\avg{k^{(A)}_{\rm in}}N}+(1-\dampingA)\frac{1}{N}.
\end{equation}
\\
\subsection*{Extension of Multiplex PageRank to multiplex networks with more than two layers.}
{
The proposed Multiplex PageRank centrality measure, presented in the main text for the specific case of a duplex network, can easily be extended to multiplex networks with more than two layers. Let us consider a multiplex network with $M$ layers given in a predetermined order, where each layer $\ell=1,2,\ldots, M$ corresponds to a network with adjacency matrix $A^{(\ell)}_{ij}$. We can define the Multiplex PageRank $X^{(\ell)}_i$ recursively in the following way. At the first level of the iteration $\ell=1$, we have the single-layer PageRank $X_i^{(1)}$ defined as
\begin{eqnarray}
X_i^{(1)} = \alpha^{(1)} \sum_{j} A_{ij}^{(1)} \frac{X_j^{(1)}}{G_j^{(1)}} + (1 - \alpha^{(1)})\frac{1}{N},
\end{eqnarray}
where $G_j^{(1)}=\max(1,\sum_r A_{rj})$. We then include the information about the structure of the other layers, and obtain
\begin{eqnarray}
X_i^{(\ell)} = \alpha^{(\ell)} \sum_j \left[X_i^{(\ell-1)}\right]^{\beta}A_{ij}^{(\ell)} \frac{X_j^{(\ell)}}{G_j^{(\ell)}} + (1 - \alpha^{(\ell)})\frac{\left[X_i^{(\ell-1)}\right]^{\gamma}}{N\Avg{\left[X^{(\ell-1)}\right]^{\gamma}}},
\label{PageRankXML}
\end{eqnarray}
where $G_j^{(\ell)}=\sum_{r} A_{rj}^{(\ell)}\left[X_r^{(\ell-1)}\right]^{\beta}+\delta\left(0,\sum_{r} \elemAdjMatA_{rj}^{(\ell)}\left[X_r^{(\ell-1)}\right]^{\beta}\right)$.
For the sake of simplicity, here we have chosen exponents $\beta$ and $\gamma$ that do not depend on $\ell$, but in general it is also possible to consider the case in which the exponents $\beta$ and $\gamma$ are dependent on the layers $\ell$. 

}

\section*{Acknowledgments}

\section*{References}



\end{document}